\documentclass[cameraready]{Interspeech}

\title{Impact Analysis of Speech Representation Learning Models for Acoustic Side-Channel Attack}

\author[affiliation={1}, equalcontribution, correspondingauthor]{Nitin}{Choudhury}
\author[affiliation={1}, equalcontribution]{Bikrant Bikram Pratap}{Maurya}
\author[affiliation={1}]{Arun Balaji}{Buduru}
\author[affiliation={2}, correspondingauthor]{Orchid Chetia}{Phukan}

\address{
    $^1$ IIIT-Delhi, India, $^2$ NTHU, Taiwan
}

\email{Correspondence: \{nitinc, orchidp\}@iiitd.ac.in}
\keywords{Acoustic Side Channel Attack, Downstream, Keystrokes, Keyboard, KAN, Pretrained Models, Representation Learning.}

\usepackage{comment}
\usepackage{amsmath}
\usepackage{multirow}
\usepackage{multicol}
\usepackage{pgfplots} 
\pgfplotsset{compat=1.17}

\begin{document}

\maketitle


\begin{abstract}

 Acoustic side-channel attacks (ASCA) on keyboards have gained increasing attention, yet impact of speech representation learning models in ASCA remains unexplored. Addressing this, we introduce \texttt{KEYAC}, a dataset designed to analyze representation generalization for ASCA under both standard and VoIP codec settings. On \texttt{KEYAC}, we evaluate six representation learning models under zero-shot and partial fine-tuning settings using fully connected and convolutional networks. Results show that while partial fine-tuning improves performance, models struggle to generalize across VoIP codecs. We hypothesize this limitation stems from inadequate modeling of nonlinear feature interactions in conventional fine-tuning architectures. To address this, we employ Kolmogorov–Arnold Networks (KAN) for fine-tuning. Empirical results show that KAN-based fine-tuning consistently outperforms the baselines and establishes a new state-of-the-art on \texttt{KEYAC}.

\end{abstract}

\section{Introduction}

 Keystroke sounds produced on a physical keyboard carry distinctive acoustic signatures~\cite{asonov2004keyboard, zhuang2009keyboard, maurya2026decker}. These signatures can be exploited to identify individual keystrokes, potentially compromising user security and privacy in real-world environments. Such attacks are commonly referred to as acoustic side-channel attacks (ASCA). In practice, users frequently type confidential information—such as passwords, personal messages, or sensitive documents—on laptops or desktops in public environments or during online meetings on platforms such as Zoom. An adversary capable of capturing these keystroke sounds may infer the typed keys and recover sensitive information.

 Although ASCA has been studied previously, existing work is constrained by several limitations. Many studies rely on datasets collected from outdated hardware or small-scale experimental setups, and the evaluations are often restricted to keyboard generalization tasks~\cite{asonov2004keyboard, zhuang2009keyboard, harrison_practical_2023, reflexnoop2024}. Moreover, most prior approaches depend on handcrafted features or spectrogram representations combined with classical machine learning (ML) or deep learning (DL) models evaluated under standard conditions~\cite{reflexnoop2024, ayati2025making}. These limitations reveal two important gaps: the absence of a public dataset that enables systematic evaluation of ASCA under VoIP codec distortions, and the lack of investigation into whether modern speech pretrained models (PTMs) can provide robust representations for ASCA tasks.

 To address these gaps, we introduce \texttt{KEYAC}, a dataset designed to analyze keystroke acoustic representations under realistic recording and transmission conditions. The dataset contains impulsive acoustic events captured using laptop microphones, smartphones, and real-time VoIP streaming pipelines. Using \texttt{KEYAC}, we evaluate six speech PTMs—Wav2Vec2~\cite{baevski2020wav2vec}, HuBERT~\cite{hsu2021hubert}, WavLM~\cite{chen2022wavlm}, OpenAI Whisper~\cite{radford2023whisper}, X-Vectors~\cite{snyder2018x}, and XLS-R~\cite{babu2021xls}—to examine their ability to generalize across keyboards and VoIP codecs. The representations extracted from these models are evaluated under zero-shot inference and lightweight downstream fine-tuning while keeping the backbone PTMs frozen, following established practices in speech representation learning~\cite{phukan2023transforming}.

 Our empirical study on \texttt{KEYAC} shows that while downstream fine-tuning improves cross-keyboard generalization, the performance of these representations deteriorates significantly under VoIP codec distortions. We hypothesize that this degradation arises from the limited ability of conventional adaptation layers to capture complex nonlinear interactions within distributed acoustic representations, particularly when spectral and temporal characteristics are altered by codec compression.

 To address this limitation, we explore a Kolmogorov–Arnold Network (KAN)~\cite{liu2024kan} based fine-tuning strategy. Instead of conventional FCN or CNN classification heads, we employ a KAN adapter to model nonlinear feature interactions more explicitly. This design is particularly suitable for scenarios where acoustic representations are affected by codec-induced spectral and temporal artifacts. Experimental results on \texttt{KEYAC} show that the proposed KAN-based adaptation consistently improves keystroke identification performance under both cross-device variability and VoIP codec distortions.

 The main contributions of this work are summarized as follows:

\begin{itemize}
    \item We introduce \texttt{KEYAC}\footnote{The dataset will me made available based on request and only for research purposes.}, a dataset designed for studying acoustic side-channel attacks under cross-device variability and VoIP codec distortions.

    \item We provide the first systematic evaluation of multiple speech pretrained models for ASCA, analyzing their generalization capability across keyboards and VoIP transmission conditions.

    \item We propose a KAN-based fine-tuning strategy and demonstrate that it consistently outperforms conventional FCN and CNN adaptation layers for keystroke identification.
\end{itemize}

\section{Background}

\subsection{Pretrained Models}

 We consider four self-supervised (SSL) speech PTM backbones for evaluation, namely, Wav2Vec2, WavLM, HuBERT, and XLS-R. Wav2Vec2 learns contextualized speech representations through a masked latent prediction framework optimized with a contrastive objective~\cite{baevski2020wav2vec}. WavLM extends this framework by jointly learning acoustic, phonetic, and speaker-level information using masked prediction combined with denoising strategies~\cite{chen2022wavlm}. HuBERT follows a similar encoder architecture to Wav2Vec2 but learns representations using a BERT-style masked prediction objective over clustered speech units~\cite{hsu2021hubert}. These models are trained in a self-supervised setting and have demonstrated strong performance across multiple speech benchmarks such as SUPERB~\cite{yang2021superb, chen2022wavlm}, SLUE~\cite{shon2022slue, shon2023slue}, and other downstream speech tasks~\cite{kodali2023classification, yang2024large, porjazovski2024raw}. 

 In addition, we include XLS-R, a multilingual extension of the Wav2Vec2 architecture designed for cross-lingual speech representation learning. XLS-R is selected due to its strong downstream performance across speech recognition, translation, and classification tasks~\cite{babu2021xls}.

 Beyond self-supervised models, we also evaluate two supervised speech representation models: X-vectors and OpenAI Whisper. X-vectors are derived from a time-delay neural network (TDNN) trained for speaker identification and are widely used for tasks such as speaker recognition (SR) and speech emotion recognition (SER)~\cite{novotny2018use, snyder2018x, phukan2023transforming}. Whisper adopts an encoder–decoder architecture trained on large-scale supervised speech data with a multitask objective. Owing to its strong zero-shot capabilities across multilingual automatic speech recognition (ASR), translation, and related speech tasks, Whisper has become a widely adopted pretrained speech model~\cite{shon2023slue, radford2023whisper, porjazovski2024raw, yang2024large, kodali2023classification}. In our experiments, we utilize only the encoder component of Whisper to extract speech embeddings.

 For feature extraction, we obtain 768-dimensional embeddings from Wav2Vec2\footnote{\url{https://huggingface.co/facebook/wav2vec2-base}}, WavLM\footnote{\url{https://huggingface.co/microsoft/wavlm-base}}, and HuBERT\footnote{\url{https://huggingface.co/facebook/hubert-base-ls960}}, 1024-dimensional embeddings from XLS-R\footnote{\url{https://huggingface.co/facebook/wav2vec2-xls-r-300m}}, and 512-dimensional embeddings from X-vectors\footnote{\url{https://huggingface.co/speechbrain/spkrec-xvect-voxceleb}} and Whisper\footnote{\url{https://huggingface.co/openai/whisper-base}}. All audio samples are resampled to a sampling rate of 16 kHz prior to feature extraction.

\subsection{Kolmogorov-Arnold Network (KAN)}

 Kolmogorov–Arnold Networks (KAN) are inspired by the Kolmogorov–Arnold representation theorem, which states that a multivariate continuous function can be expressed as a composition of simpler one–dimensional functions~\cite{liu2024kan}. In other words, complex interactions between multiple variables can be approximated by combining a set of learnable univariate transformations. Formally, a function \(f(x_1,\dots,x_n)\) can be expressed as a sum of functions operating on transformed inputs, $f(x_1,\dots,x_n) = \sum_{q} \Phi_q \left(\sum_{p} \psi_{q,p}(x_p)\right)$, where \(\psi_{q,p}(\cdot)\) and \(\Phi_q(\cdot)\) denote learnable one–dimensional functions.

 Based on this idea, KAN differs from conventional neural networks by replacing fixed linear weights with learnable nonlinear functions along the network connections. Given an input vector \(x=(x_1,\dots,x_n)\), a KAN layer aggregates transformed inputs as $h_j = \sum_{i} \phi_{i,j}(x_i)$, where \(\phi_{i,j}(\cdot)\) represents a learnable nonlinear mapping between the input \(x_i\) and hidden unit \(j\). The final output is obtained by combining these intermediate representations using another set of learnable functions.

 In practice, these functions are typically implemented using spline bases (e.g., B-splines), which allow flexible yet smooth function approximation. This design enables KAN to model nonlinear feature interactions more explicitly than standard multilayer perceptrons that rely on fixed activation functions and scalar weights. Recent studies have shown that KAN can be effective across several domains, including function approximation, tabular prediction, scientific modeling, and parameter-efficient adaptation of pretrained models~\cite{liu2024kan, cai2025loki, li2024kan}.

\section{Methodology}

\subsection{Dataset}

Existing Acoustic Side-Channel Attack (ASCA) datasets are often limited by outdated hardware, small-scale collections, or evaluations restricted to basic keyboard generalization tasks. Moreover, prior datasets lack the diversity needed to study representation stability across modern telecommunication channels. To address these limitations, we introduce \texttt{KEYAC}, a dataset designed to analyze ASCA representation generalization under both standard acoustic conditions and Voice-over-IP (VoIP) codec settings. \texttt{KEYAC} consists of impulsive acoustic events (keystrokes) captured across three recording channels: local laptop microphones, smartphones (near-field), and real-time VoIP streaming pipelines such as Zoom and Teams. The dataset contains recordings from 37 keyboards with a total of 37,440 keystroke events, distributed evenly across the three channels, each contributing 12,480 samples. These channels represent distinct acoustic environments including line-of-sight (LoS), non-line-of-sight (NLoS), and codec-distorted conditions introduced by automatic gain control (AGC) and acoustic echo cancellation (AEC). This multi-channel design enables systematic evaluation of cross-device variability and VoIP-induced representation drift, providing a practical benchmark for studying speech pretrained models and downstream adaptation strategies in realistic ASCA scenarios.

\subsection{Baseline Downstream Modeling of Individual Backbones}
\label{subsec:downstream}

\begin{figure}
    \centering
    \includegraphics[width=0.89\linewidth]{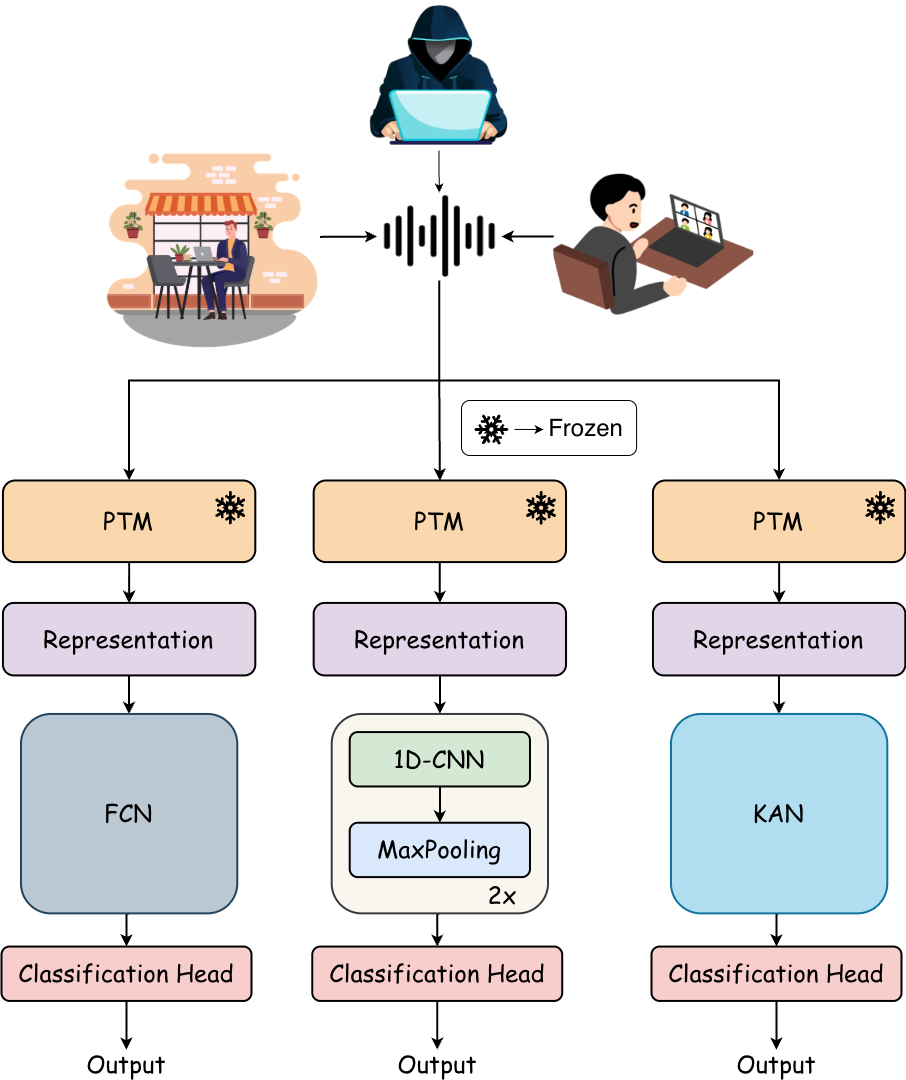}
    \caption{Downstream Architectures Considered for Evaluation}
    \label{fig:placeholder}
\end{figure}

For experimentation, we consider two downstream strategies following previous research~\cite{phukan2025rethinking, choudhury2026foca, choudhury2026gocoma}--a fully connected network (FCN) and a convolutional neural network (CNN) as shown in figure~\ref{fig:placeholder}. These models are evaluated independently for both detection and fraud-type characterization. The downstream architectures are intentionally kept simple to better assess the intrinsic capability of the PTM representations for cross-task evaluation.

 For the FCN, we employ a dense layer with 90 hidden neurons followed by the classification layer, using sigmoid and softmax activations for multiclass classification tasks, respectively. For the CNN-based downstream, we use two convolutional blocks, each consisting of a convolutional layer with kernel size 3 followed by a max-pooling layer with pool size 2. The output from the final convolution block is then passed to the same classification layer as used in the FCN.

 \textit{\textbf{Training.}} We utilize categorical cross-entropy losses for the multiclass classification task. The models are optimized using AdamW and trained for 20 epochs with a batch size of 16 and a learning rate of 2.2e-4. Dropout and early stopping are used to reduce overfitting.

\subsection{KAN-based Downstream Modeling}

To further examine whether nonlinear feature interactions can improve the utilization of pretrained representations, we consider a KAN-based downstream model. In this setup, the conventional linear transformations are replaced with learnable univariate spline functions following the Kolmogorov–Arnold network formulation. The model consists of a single hidden KAN layer with 30 units, considering grid to be 5, and spline order to be 3, followed by a task-specific output layer. As with the baseline downstream models, softmax activation is used for calculating the multiclass probability, followed by applying argmax for final classification.

\subsection{Experimentation Strategy}

The pretrained model (PTM) encoders are kept frozen and only the downstream models are trained. We evaluate the learned representations under both in-domain (ID) and out-of-domain (OOD) settings.

For the in-domain (ID) setup, we follow a 5-fold cross-validation protocol where four folds are used for training and one fold is used for testing in each iteration. For out-of-domain evaluation, we consider three generalization scenarios, each defined by a fixed train-test channel configuration, with ID and OOD differing only in keyboard exposure.

First, we evaluate keyboard generalization under standard recording conditions using a keyboard holdout strategy. In this setup, samples from one keyboard are held out for testing while the models are trained on samples from the remaining keyboards, both recorded in the standard format. This setting evaluates the ability of the models to generalize to unseen keyboard hardware.

Second, we evaluate keyboard generalization under VoIP codec conditions. The keyboard holdout protocol is retained, but both training and evaluation are performed entirely on VoIP-streamed audio. This setup isolates the effect of keyboard variability alone under a fixed VoIP channel.

Third, we evaluate codec generalization independently of keyboard variation. In this setting, the models are trained on standard recordings and evaluated on VoIP-streamed audio samples, in both the ID and OOD settings. This protocol isolates the effect of codec-induced spectral and temporal distortions on ASCA performance.






\section{Results and Discussion}

\begin{table*}[t]
\centering
\caption{Generalization performance on KEYAC. Results reported as Accuracy (\%) / Macro-F1 (\%).}
\begin{tabular}{lcccccccccccc}
\toprule
\multirow{3}{*}{Models} &
\multicolumn{4}{c}{Keyboard Generalization (Standard)} &
\multicolumn{4}{c}{Keyboard Generalization (VoIP)} &
\multicolumn{4}{c}{Codec Generalization} \\

\cmidrule(lr){2-5}
\cmidrule(lr){6-9}
\cmidrule(lr){10-13}

& \multicolumn{2}{c}{ID} & \multicolumn{2}{c}{OOD}
& \multicolumn{2}{c}{ID} & \multicolumn{2}{c}{OOD}
& \multicolumn{2}{c}{ID} & \multicolumn{2}{c}{OOD} \\

\cmidrule(lr){2-3} \cmidrule(lr){4-5}
\cmidrule(lr){6-7} \cmidrule(lr){8-9}
\cmidrule(lr){10-11} \cmidrule(lr){12-13}

& Acc & mF1 & Acc & mF1 & Acc & mF1 & Acc & mF1 & Acc & mF1 & Acc & mF1 \\

\midrule
\multicolumn{13}{c}{\textbf{CNN Downstream}} \\
\midrule
Wav2Vec2 & 47.82 & 46.41 & 38.95 & 37.64 & 33.71 & 32.26 & 25.14 & 23.88 & 34.15 & 32.78 & 26.54 & 25.18 \\
WavLM    & \textbf{58.34} & \textbf{56.92} & 49.47 & 48.16 & \textbf{44.12} & \textbf{42.73} & 35.86 & 34.54 & \textbf{45.21} & \textbf{43.86} & 37.54 & 36.21 \\
HuBERT   & 36.27 & 34.84 & 27.61 & 26.18 & 25.92 & 24.71 & 18.66 & 17.42 & 26.83 & 25.52 & 19.44 & 18.21 \\
XLS-R    & 33.45 & 32.08 & 24.82 & 23.47 & 23.41 & 22.24 & 16.75 & 15.58 & 24.12 & 22.91 & 17.36 & 16.04 \\
X-Vectors& 44.18 & 42.86 & 35.74 & 34.21 & 30.27 & 28.91 & 22.65 & 21.36 & 31.28 & 29.91 & 23.76 & 22.41 \\
Whisper  & 48.73 & 47.41 & 39.96 & 38.54 & 35.64 & 34.31 & 27.54 & 26.21 & 36.47 & 35.14 & 28.62 & 27.23 \\

\midrule
\multicolumn{13}{c}{\textbf{FCN Downstream}} \\
\midrule
Wav2Vec2 & 45.36 & 43.92 & 36.54 & 35.11 & 31.68 & 30.24 & 23.94 & 22.71 & 32.47 & 31.08 & 24.82 & 23.54 \\
WavLM    & 55.81 & 54.36 & 46.94 & 45.52 & 40.35 & 39.02 & 32.48 & 31.14 & 41.72 & 40.41 & 33.86 & 32.44 \\
HuBERT   & 34.41 & 33.08 & 26.02 & 24.78 & 23.85 & 22.64 & 17.21 & 16.02 & 24.73 & 23.51 & 17.92 & 16.67 \\
XLS-R    & 31.74 & 30.36 & 23.42 & 22.11 & 21.64 & 20.43 & 15.02 & 13.88 & 22.56 & 21.27 & 15.94 & 14.72 \\
X-Vectors& 41.63 & 40.18 & 33.21 & 31.82 & 28.26 & 26.91 & 20.73 & 19.41 & 29.54 & 28.16 & 22.11 & 20.74 \\
Whisper  & 46.25 & 44.91 & 37.58 & 36.17 & 33.14 & 31.86 & 25.27 & 23.96 & 34.08 & 32.76 & 26.42 & 25.07 \\

\midrule
\multicolumn{13}{c}{\textbf{KAN Downstream (Proposed)}} \\
\midrule
Wav2Vec2 & 66.18 & 64.82 & 59.74 & 58.33 & 55.91 & 54.62 & 49.36 & 48.02 & 56.72 & 55.41 & 51.28 & 49.93 \\
WavLM    & \textbf{68.47} & \textbf{67.12} & 61.73 & 60.34 & \textbf{57.82} & \textbf{56.41} & 50.91 & 49.56 & \textbf{58.36} & \textbf{57.04} & 52.84 & 51.46 \\
HuBERT   & 63.28 & 61.94 & 56.47 & 55.11 & 52.06 & 50.73 & 46.23 & 44.91 & 52.18 & 50.86 & 47.21 & 45.83 \\
XLS-R    & 61.74 & 60.42 & 54.93 & 53.62 & 50.47 & 49.12 & 44.51 & 43.22 & 50.96 & 49.64 & 46.08 & 44.71 \\
X-Vectors& 58.41 & 57.06 & 51.62 & 50.23 & 47.18 & 45.86 & 41.63 & 40.32 & 48.17 & 46.83 & 42.96 & 41.52 \\
Whisper  & 65.74 & 64.39 & 59.12 & 57.83 & 54.83 & 53.46 & 48.92 & 47.63 & 55.63 & 54.28 & 50.41 & 49.06 \\

\bottomrule

\label{tab:results}
\end{tabular}
\end{table*}

 We evaluate the models using accuracy and macro-F1 (mF1) to analyze their effectiveness across three generalization scenarios: keyboard generalization under standard recordings, keyboard generalization under VoIP conditions, and codec generalization. The results are summarized in Table~\ref{tab:results}. Across all settings, the PTMs exhibit different levels of robustness depending on both the downstream architecture and the evaluation scenario.

 For keyboard generalization under standard recording conditions, the CNN downstream generally provides the strongest baseline performance. Among the PTMs, WavLM consistently achieves the highest performance. For instance, with the CNN downstream, WavLM achieves 58.34\% accuracy and 56.92\% mF1 in the in-domain setting, outperforming other PTMs by a noticeable margin. In contrast, models such as HuBERT and XLS-R exhibit substantially lower performance, with in-domain accuracy values of 36.27\% and 33.45\%, respectively. This gap indicates that while several speech PTMs capture useful acoustic information, WavLM representations appear more suitable for modeling impulsive keyboard sounds. Under the keyboard holdout (OOD) setting, performance decreases across all models by roughly 8–10\%, indicating the challenge of generalizing to unseen keyboard hardware.

 When evaluating keyboard generalization under VoIP conditions, the overall performance further decreases due to codec-induced distortions. Again, WavLM remains the most robust backbone among the evaluated PTMs. With the CNN downstream, WavLM achieves 44.12\% accuracy and 42.73\% mF1 in the in-domain setting, while other models experience a larger performance drop. HuBERT and XLS-R, for example, obtain only 25.92\% and 23.41\% accuracy respectively. The OOD results reveal an additional degradation of approximately 8–10\%, reflecting the combined difficulty of keyboard variation and codec compression artifacts.

 For the codec generalization task, where models are trained on standard recordings and evaluated on VoIP-streamed audio, the performance gap across PTMs becomes even more evident. WavLM again achieves the best baseline performance with the CNN downstream, reaching 45.21\% accuracy and 43.86\% mF1. Other PTMs show significantly lower scores, confirming that codec distortions substantially affect the discriminative cues required for keystroke identification.

 Across all three tasks, the choice of downstream architecture plays an important role. While CNN generally performs slightly better than FCN among the baseline architectures, both approaches exhibit limited capability in modeling the complex nonlinear interactions present in the pretrained acoustic representations. This limitation becomes particularly evident under VoIP conditions, where spectral and temporal distortions significantly alter the acoustic characteristics of keystrokes.

 In contrast, the proposed KAN-based downstream consistently improves performance across all PTMs and evaluation scenarios. For keyboard generalization under standard recordings, KAN achieves the best overall performance with the WavLM backbone, reaching 68.47\% accuracy and 67.12\% mF1 in the in-domain setting. Even under the keyboard holdout scenario, WavLM with KAN maintains strong performance with 61.73\% accuracy and 60.34\% mF1, representing a substantial improvement over both CNN and FCN baselines.

 A similar trend is observed under VoIP conditions. With KAN, WavLM achieves 57.82\% accuracy and 56.41\% mF1 for keyboard generalization under VoIP recordings, significantly outperforming the CNN baseline. For the codec generalization task, the KAN-based downstream again yields the best results, achieving 58.36\% accuracy and 57.04\% mF1 with the WavLM backbone. Even in the OOD codec setting, the performance degradation remains relatively moderate compared to the baseline downstream models.

 Overall, these results highlight two key observations. First, among the evaluated PTMs, WavLM provides the most robust acoustic representations for keystroke identification across both keyboard and codec variations. Second, the proposed KAN-based downstream consistently improves performance by effectively modeling nonlinear feature interactions within the pretrained representations. This advantage becomes particularly important under VoIP codec distortions, where conventional linear or shallow nonlinear adaptation layers struggle to capture the altered acoustic patterns. These findings suggest that combining robust speech representations with expressive nonlinear adaptation layers is critical for improving ASCA generalization under realistic recording and transmission conditions.

\section{Conclusion}

 In this work, we introduced \texttt{KEYAC}, a dataset designed to study acoustic side-channel attacks under realistic conditions, including cross-device variability and VoIP codec distortions. Using this dataset, we systematically evaluated multiple speech pretrained models to examine their generalization capability for keystroke identification across keyboards and codec environments. Our results show that although modern speech representations provide useful features for ASCA, their performance degrades under VoIP compression and unseen keyboard conditions when conventional FCN or CNN downstream architectures are used. To address this limitation, we explored a Kolmogorov–Arnold Network (KAN) based downstream adaptation strategy that models nonlinear feature interactions more effectively. Experimental results demonstrate that the KAN-based downstream consistently improves performance across all PTMs and generalization scenarios, with WavLM achieving the strongest results. These findings highlight the importance of expressive downstream modeling for improving ASCA robustness under realistic recording and transmission conditions.

\section{Generative AI Use Disclosure}
Generative AI has been used only for linguistic editing and improving readability, and not for ideation, methodological contribution, results, or content development.

\bibliographystyle{IEEEtran}
\bibliography{mybib}

\end{document}